\documentclass[iop, twocolappendix, tighten]{emulateapj}
\usepackage{apjfonts}
\usepackage{verbatim}
\usepackage{color}
\usepackage{natbib}
\usepackage{amssymb,amsmath,mathrsfs}
\tabletypesize{\scriptsize}

\usepackage{hyperref}
\definecolor{linkcolor}{rgb}{0,0,0.5}
\hypersetup{colorlinks=true,linkcolor=linkcolor,citecolor=linkcolor,
           filecolor=linkcolor,urlcolor=linkcolor}
\usepackage{url}
\usepackage{amssymb,amsmath}
\usepackage{subfigure}

\def\h2{$\rm H_2$}

\def\halpha{H$\alpha$}

\newcommand{\bvec}[1]{\ensuremath{\boldsymbol{#1}}}

\newcommand{\msun}{M$_{\odot}$}
\newcommand{\zsun}{Z$_{\odot}$}

\newcommand{\thetamath}{\ensuremath{\bvec{\theta}}}

\newcommand{\Qk}{\ensuremath{\bvec{Q}}}

\newcommand{\gammamath}{\ensuremath{\bvec{\Gamma}}}

\begin{document}

\title{The High-Mass Stellar Initial Mass Function in M31 Clusters\altaffilmark{*}}

\author{
Daniel R.\ Weisz\altaffilmark{1,12},
L.\ Clifton Johnson\altaffilmark{1},
Daniel Foreman-Mackey\altaffilmark{2},
Andrew E.\ Dolphin\altaffilmark{3},
Lori C.\ Beerman\altaffilmark{1},
Benjamin F.\ Williams\altaffilmark{1},
Julianne J.\ Dalcanton\altaffilmark{1},
Hans-Walter Rix\altaffilmark{4},
David W.\ Hogg\altaffilmark{2,4},
Morgan Fouesneau\altaffilmark{4},
Benjamin D.\ Johnson\altaffilmark{5},
Eric F.\ Bell\altaffilmark{6},
Martha L.\ Boyer\altaffilmark{7},
Dimitrios Gouliermis\altaffilmark{4,8},
Puragra Guhathakurta\altaffilmark{9},
Jason S.\ Kalirai\altaffilmark{7},
Alexia R.\ Lewis\altaffilmark{1},
Anil C.\ Seth\altaffilmark{10},
Evan D.\ Skillman\altaffilmark{11}
}

\altaffiltext{*}{Based on observations made with the NASA/ESA Hubble Space Telescope, 
obtained at the Space Telescope Science Institute, which is operated by the 
Association of Universities for Research in Astronomy, Inc., under NASA contract NAS 
5-26555. These observations are associated with program \#12055}
\altaffiltext{1}{Astronomy Department, Box 351580, University of Washington, Seattle, WA, USA; 
dweisz@uw.edu}
\altaffiltext{2}{Center for Cosmology and Particle Physics, New York, University, 4 Washington Place, New 
York, NY 10003, USA}
\altaffiltext{3}{Raytheon Company, 1151 East Hermans Rd, Tucson, AZ 85756, USA}
\altaffiltext{4}{Max Planck Institute for Astronomy, Koenigstuhl 17, 69117 Heidelberg, Germany}
\altaffiltext{5}{Harvard-Smithsonian Center for Astrophysics, 60 Garden Street, Cambridge, MA 02138, USA}
\altaffiltext{6}{Department of Astronomy, University of Michigan, 500 Church Street, Ann Arbor, MI 48109, USA}
\altaffiltext{7}{Space Telescope Science Institute, 3700 San Martin Drive, Baltimore, MD 21218, USA}
\altaffiltext{8}{Universit\"at Heidelberg, Zentrum f\"ur Astronomie, Institut f\"ur Theoretische Astrophysik,
Albert-Ueberle-Str.~2, 69120 Heidelberg, Germany}
\altaffiltext{9}{Department of Astronomy, University of California at Santa Cruz,
1156 High Street, Santa Cruz, CA, 95064, USA}
\altaffiltext{10}{Department of Physics and Astronomy, University of Utah,Salt Lake City, UT 84112, USA}
\altaffiltext{11}{Minnesota Institute for Astrophysics, University of Minnesota, 116 Church Street SE, Minneapolis, MN 55455, USA}
\altaffiltext{12}{Hubble Fellow}


\begin{abstract}

We have undertaken the largest systematic study of the high-mass stellar initial mass function (IMF) to date using the optical color-magnitude diagrams (CMDs) of 85 resolved, young ($\rm 4~Myr < t <25~Myr$), intermediate mass star clusters (10$^3$-10$^4$ \msun), observed as part of the Panchromatic Hubble Andromeda Treasury (PHAT) program.  We fit each cluster's CMD to measure its mass function (MF) slope for stars $\ga$ 2 \msun. By modeling the ensemble of clusters, we find the distribution of MF slopes is best described by $\Gamma=+1.45^{+0.03}_{-0.06}$ with a very small intrinsic scatter.   This model allows the MF slope to depend on cluster mass, size, and age, but the data imply no significant dependencies within this regime of cluster properties. The lack of an age dependence suggests that the MF slope has 
 not significantly evolved over the first $\sim 25$~Myr, and provides direct observational 
 evidence that the measured MF represents the IMF.
 Taken together, this analysis --- based on an unprecedented large sample of 
 young clusters, homogeneously constructed CMDs, well-defined selection 
 criteria, and consistent principled modeling --- implies that the high-mass IMF slope in M31 clusters is universal.  The IMF has a slope ($\Gamma=+1.45^{+0.03}_{-0.06}$) that is slightly steeper than the canonical Kroupa ($+1.30$) and Salpeter ($+1.35$) values, with no drastic outliers in this sample of nearly 100 clusters. Using our inference model on select Milky Way (MW) and LMC high-mass IMF studies from the literature, we find $\Gamma_{\rm MW} \sim+1.15\pm0.1$ and $\Gamma_{\rm LMC} \sim+1.3\pm0.1$, both with intrinsic scatter of $\sim$0.3--0.4 dex.  Thus, while the high-mass IMF in the Local Group may be universal,  systematics in literature IMF studies preclude any definitive conclusions; homogenous investigations of the high-mass IMF in the local universe are needed to overcome this limitation. \emph{Consequently, the present study represents the most robust measurement of the high-mass IMF slope to date}.  To facilitate practical use over the full stellar mass spectrum, we have grafted the M31 high-mass IMF slope onto widely used sub-solar mass Kroupa and Chabrier IMFs.  The increased steepness in the M31 high-mass IMF slope implies that commonly used UV- and \halpha-based star formation rates should be increased by a factor of $\sim$1.3--1.5  and the number of stars with masses $>8$ \msun\ are $\sim 25$\% fewer than expected for a Salpeter/Kroupa IMF.

\end{abstract}



\section{Introduction}
\label{sec:intro}

The stellar initial mass function (IMF), for $M\ge $ 1 \msun\ plays a central role in a wide 
variety of astrophysics.  The relative numbers of such stars is 
essential to interpreting the stellar populations of star-forming galaxies across 
cosmic time, testing and validating theories of star 
formation, constraining chemical evolution models, the formation of compact objects and gravitational waves, and understanding the interplay between stars and gas \citep[e.g.,][]{schmidt1959, tinsley1968, searle1973, talbot1973, ostriker1974, audouze1976, scalo1986, kennicutt1998b, elmegreen1999, kroupa2001, massey2003, alves2007, fardal2007, mckee2007, wilkins2008, pflammAltenburg2009, kennicutt2012, narayanan2012, conroy2013, belczynski2014, krumholz2014, madau2014b}.

\subsection{Current State of the IMF Above $\sim$ 1 \msun}
\label{sec:currentimf}

Our best constraints on the high-mass IMF come from studies in the Milky Way (MW) and nearby galaxies.  For the closest galaxies, it is possible to count individual stars and make a direct measurement of the high-mass IMF slope.  While measuring the IMF from direct star counts is straightforward in principle, it is far more complicated in practice.  Accurate measurements are challenging due to a variety of observational and physical effects including completeness, dynamical evolution, distance/extinction ambiguity, stellar multiplicity, the accuracy of stellar evolution models, degeneracies between star formation history (SFH) and the IMF slope, and a paucity of massive stars in the nearby universe, all of which complicate the translation of the present day mass function (MF) into the IMF \citep[e.g.,][]{lequeux1979, miller1979, scalo1986, mateo1993, massey1995, delafuentemarcos1997, maizapellaniz2005, elmegreen2006, maizapellaniz2008, ascenso2009, kruijssen2009, demarchi2010, portegieszwart2010, bastian2010}.  

Observations of more distant, unresolved galaxies alleviate some of these challenges.  For example, by studying an entire galaxy, effects of stellar dynamics (e.g., migration, mass segregation) and distance/extinction confusion are less important.  Further, distant galaxies span a more diverse range of environments than what is available in the very local universe, offering better opportunities to measure the high-mass IMF as a function of environment \citep[metallicity, mass, redshift, etc.; e.g.,][]{baldry2003, fardal2007, dave2011, narayanan2013}.  However, the unresolved nature of these observations severely limits the precision and accuracy of a high-mass IMF measurement.  Most integrated observations have insufficient leverage to mitigate the SFH-IMF degeneracy, and are forced to make simplified assumptions about a galaxy's recent SFH (e.g., constant, single short burst) and dust (e.g., a single value for an entire galaxy) to provide \emph{any} constraint on the high-mass IMF \citep[e.g.,][]{miller1979, elmegreen2006}.  Although studies of unresolved clusters may provide a new avenue for IMF studies \citep[e.g.,][]{calzetti2010, andrews2013}, high-mass IMF studies using integrated light typically provide coarse characterizations rather than precise measurements.

As a result of these challenges, our current knowledge of the high-mass IMF remains remarkably poor. The widespread adoption of a `universal' Salpeter \citep[$\Gamma=+1.35$;][]{salpeter1955} or Kroupa \citep[$\Gamma=+1.3$;][]{kroupa2001} IMF above 1 \msun, presumed not to vary with environment has given the impression that the high-mass IMF is a solved problem \citep[e.g.,][]{kennicutt1998b}.  However, compilations of literature IMF studies from resolved star counts indicate an ``average value'' of $\Gamma=1.7$ with a scatter $\gtrsim$0.5 dex in IMF slope measurements that do not clearly support a `universal' IMF \citep[e.g.,][]{scalo1986, scalo1998, kroupa2002, scalo2005}\footnote{Although a `Kroupa IMF' has a high-mass slope of $\Gamma=1.3$, \citet{kroupa2002} argue that the slope should probably be close $\Gamma \sim 1.7$, as most historical IMF studies do not account for unresolved binary stars, which, in principle, cause measured MF slopes to be shallower than true IMF slopes.}.  There also does not appear to be any unambiguous systematic dependence of IMF slope on environment and/or initial star-forming conditions \citep[e.g.,][]{bastian2010}, which has significant implications for how stars form in a wide variety of physical conditions \citep[e.g.,][]{elmegreen1999, mckee2007, hopkins2013a, krumholz2014}.

Controversy over the form and universality of the IMF have persisted since the first studies in the 1950s.
Following the seminal work of \citet{salpeter1955}, a handful of studies reported
similar IMFs in a number of MW clusters and field environments \citep{jaschek1957, sandage1957, vandenbergh1957}, prompting early speculation about the universality of the ``original mass function''  \citep[e.g.,][]{limber1960, vandenbergh1960}.  However, this early agreement with Salpeter gave way to large scatter in IMF measurements beginning in the 1970s, as several studies with more sophisticated completeness corrections and updated stellar physics highlighted the shortcomings in the first generation of IMF studies \citep[e.g.,][]{taff1974,lequeux1979, tarrab1982}.  Furthermore, \citet{miller1979} provided a comprehensive reassessment of the MW field star IMF, in light of the previously unaccounted for SFH-IMF degeneracy.  They suggested a high-mass IMF of the with $\Gamma=1.5$ for stars $1\lesssim$M/\msun$\lesssim$10 and $\Gamma=2.3$ for stars M/\msun$\ga$10.  However, \citet{kennicutt1983} showed that such a steep IMF slope above $\sim$10 \msun\ was incompatible with observations of $B-V$ colors and \halpha\ equivalent widths in nearby disk galaxies, suggesting that extrapolating the value of $\Gamma=1.5$ from 1-100 \msun\ provided a better match to the data.  This conclusion was further solidified in \citet{kennicutt1994}.

The number of observational high-mass IMF studies peaked between the late 1980s and early 2000s.  This golden age followed the comprehensive IMF review by \citet{scalo1986}, which concluded that systematic uncertainties were the dominant source of confusion in our poor knowledge of the IMF.  Shortly after, the proliferation of CCDs enabled significant progress in reducing these systematics, primarily by allowing for quantitative completeness corrections via artificial star tests \citep[e.g.,][]{stetson1987, mateo1993}.  The 1990s saw a flurry of qualitatively new IMF studies including star counts in the LMC, spectroscopy of starburst galaxies, and IMF constraints from stellar abundance patterns (see \citealt{leitherer1998, massey1998, gilmore2001} for extensive discussion of the literature in these areas).  The sheer number and richness of IMF studies from this era precludes a detailed history in this paper, and we instead refer readers to a number of excellent papers that chronicle the history of the IMF \citep[e.g.,][]{scalo1986, massey1998, kennicutt1998b, elmegreen1999, eisenhauer2001, gilmore2001, massey2003, scalo2005, elmegreen2006b, zinnecker2005, bastian2010}.

To briefly summarize, many studies from this era concluded that the high-mass IMF in the local universe was not drastically different from Salpeter over a wide range of environments \citep[e.g.,][]{massey1998, leitherer1998, gilmore2001, kroupa2001, massey2003}.  However, there was no clear explanation for the observed scatter in the measured IMF slopes, which contradicts a truly universal high-mass IMF.  Some have argued that measurement uncertainty and systematics are the dominant source of scatter \citep[e.g.,][]{elmegreen1999, kroupa2001, kroupa2002, massey2003}. However, without systematically re-visiting each study, it is challenging to dismiss the scatter as solely a measurement uncertainty.  As a clear counter example, there are well-known studies of star clusters with some of the same physical properties (e.g., age, mass) that have been analyzed identically, but have very different high-mass IMF slopes \citep[e.g.,][]{phelps1993}, reinforcing that some of the scatter may be physical in nature.  At the same time, on the whole, $>$75\% of resolved star IMF studies have significantly underreported uncertainties relative to the theoretically possible precision \citep{weisz2013a}, which underlines the difficulty in disentangling systematics from intrinsic IMF variation using literature IMF studies.

Since the early 2000s, studies of the high-mass IMF have diminished in number, as attention turned to the sub-solar regime \citep[e.g.,][]{chabrier2003, scalo2005, zinnecker2005}.  However, focus on the high-mass IMF was recently rekindled following several reports of a systematic deficit of massive stars in lower-luminosity galaxies, based on observations of integrated broadband and/or \halpha-to-UV flux ratios \citep[e.g.,][]{hoversten2008, boselli2009, lee2009, meurer2009}.  Yet, whether these observations are indicative of a systematically varying IMF \citep[e.g.,][]{weidner2005, pflammAltenburg2009} remains an open question, as other mechanisms such as stochastic sampling of the cluster mass function and/or bursty SFHs can also explain these observed properties \citep[e.g.,][]{fumagalli2011, dasilva2012, weisz2012a}. 

Thus, at present, the high-mass IMF is characterized by common usage of a universal Salpeter/Kroupa IMF, but also boasts a vast body of literature studies which does not unambiguously support this as the \textit{de facto} standard.  After decades of study, the main limitation in our knowledge of the high-mass IMF remains our reliance on a rich, but heterogenous set of literature IMF studies, instead of a large, homogenous survey of individually resolved young, massive stars.

\subsection{The IMF $>$ 2--3 \msun\ in M31}
\label{sec:phat_imf}

A primary goal of the Panchromatic Hubble Andromeda Treasury program \citep[PHAT;][]{dalcanton2012} is a comprehensive exploration of the stellar IMF above $\sim$ 2-3 \msun. This 828-orbit \textit{HST} multi-cycle program acquired near-UV through the near-IR imaging of $\sim$120 million 
resolved stars in $\sim$ 1/3 of M31's star-forming disk \citep{williams2014}, providing a dataset of unprecedented size and homogeneity for measuring the high-mass IMF.

The focus of the present IMF study is on M31's  young star cluster population.  Star clusters provide a more direct path to measuring the IMF compared to field populations, which suffer from IMF-SFH degeneracies for stars more luminous than the oldest main sequence turnoff \citep[e.g.,][]{miller1979, elmegreen2006}, Further, the differential extinction found in the disk of M31 \citep[e.g.,][]{draine2013, dalcanton2015}, and similarly sized star-forming galaxies, adds a significant layer of complexity to modeling the field stellar populations.  In contrast, star clusters provide a relatively simple way to probe the IMF.  Their co-eval nature genuinely mitigates the effects of SFH, while their vulnerability to foreground differential extinction is minimal due to their small sizes.

In this paper, we focus on determining the MF slopes for PHAT clusters that are $\la 25$ Myr old and $\ga10^3$ \msun\ in mass using resolved star color-magnitude diagrams (CMDs).  For these clusters, the detectable portion of the main sequence is well-populated from $\sim$ 2\msun\ to $\ga 25$\msun\ and they are of order $\sim$ 0.1 relaxation times old, which mitigates the strongest effects of mass segregation, making them an ideal sample with which to study the high-mass IMF.  We model the optical CMD of each cluster $k$ to constrain its power-law slope $\Gamma_k$ of the present day high-mass MF, ${\mathrm p}(\Gamma_k)$, marginalized over other parameters of that fit (e.g., cluster age, extinction, binarity).  

Subsequently, we propose a simple model for the ensemble properties of the cluster MF, with a mean, $\Gamma$, and a scatter, $\sigma_\Gamma$, and fit it to the MF probability distribution functions (PDFs) of the individual clusters.  We also use this model to search for possible dependencies of $\Gamma$ on cluster mass, age, and size.  We verify this approach with an extensive set of artificial clusters that were inserted into PHAT images, processed, and analyzed identically to the real clusters.

This paper is organized as follows.  In \S \ref{sec:data}, we define our cluster sample and briefly describe the PHAT photometry, artificial star tests, and artificial cluster sample.  We outline our method of modeling the cluster CMDs in \S \ref{sec:methodology} and present the results for the ensembles of real and artificial clusters in \S \ref{sec:results}. Finally, in \S \ref{sec:discussion}, we discuss our findings in the context of current and historical views on the high-mass IMF and illustrate the implications of our results

\section{Data}
\label{sec:data}

\subsection{Cluster Sample}
\label{sec:sample}

Clusters in PHAT were identified as part of the Andromeda Project\footnote{\url{http://www.andromedaproject.org}}, a collaborative effort between PHAT and the Zooniverse\footnote{\url{https://www.zooniverse.org}} citizen science platform.  As described in \citet{johnson2015}, cluster candidates were assigned probabilities of actually being a cluster based on citizen scientist ranking.  These probabilities were verified using synthetic cluster results and comparison with the preliminary catalog, which was assembled by professional astronomers \citep{johnson2012}.  Extensive details of the cluster identification methodology can be found in \citet{johnson2015}.  

For the present study, we have limited our analysis to the 85 clusters with CMD-based ages $\lesssim 25$ Myr and masses $\gtrsim 10^3$ \msun\ \citep{beerman2015}, and half-light radii $\ga$ 2 pc \citep{johnson2015}. We focus on these young, more massive clusters in PHAT to avoid the very sparse sampling in the low-mass clusters; to limit the impact of stellar crowding in the densest clusters; to facilitate the extrapolation of the observable MF to an IMF; and to remain in the regime of high completeness ($>$90\% for our sample; see Figure 10 in \citealt{johnson2015}).  We will explore each of these issues in detail in our comprehensive study of PHAT cluster MFs in a future paper.

\subsection{Cluster Photometry}
\label{sec:photometry}

\begin{figure}[t!] 
\epsscale{0.85}
 \plotone{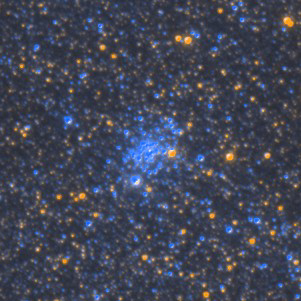} 
\caption{The HST/ACS optical color image of AP~94, a typical cluster analyzed in this paper.  The angular high-resolution of HST allows us to resolve the individual stars clusters at the distance of M31.  For reference, the image is $\sim$50pc on each side.}
\label{fig:image}
\end{figure}

\begin{figure}[t!] 
\epsscale{1.2}
 \plotone{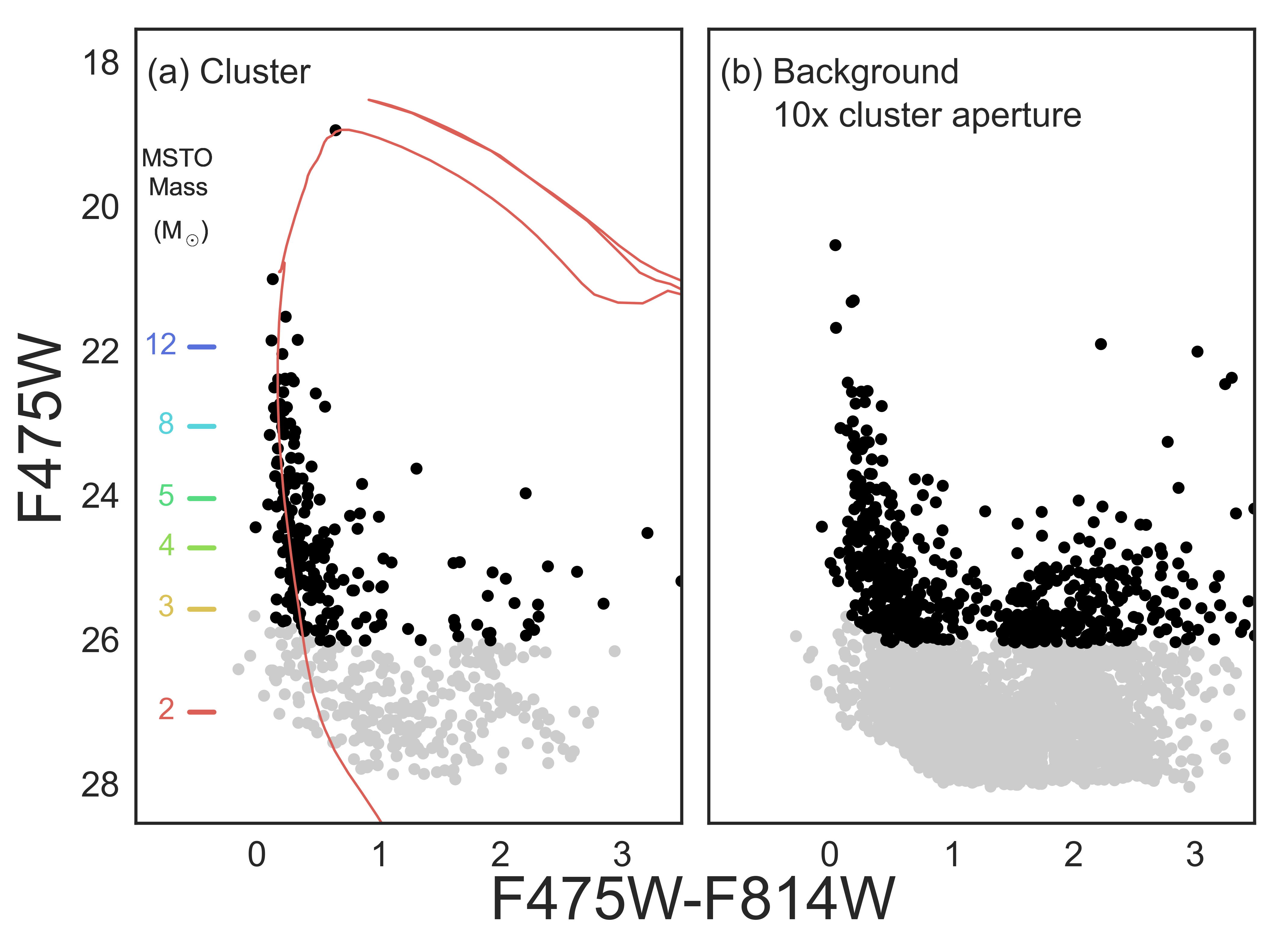} 
\caption{Panel (a) -- The optical CMD of AP 94.  The grey points are below the nominal completeness limit, and were not included in the CMD analysis.  For reference, we indicate the main sequence turn off masses on the left hand side and over-plot the best fitting isochrone for this cluster (see \S \ref{sec:match} and Figure \ref{fig:triangle_match}).  Panel (b) -- the background CMD constructed by selecting all stars in an annulus surrounding the cluster that has 10x times the cluster area.  The black points in this CMD are used to statistically model background contribution to the cluster CMD.}
\label{fig:cmd}
\end{figure}

Photometry of PHAT clusters was performed using DOLPHOT\footnote{\url{http://americano.dolphinsim.com}}, a version of \texttt{HSTPHOT} \citep{dolphin2000b} with HST specific modules, following the description in \citet{williams2014}\footnote{\url{http://archive.stsci.edu/missions/hlsp/phat/}}.  Slight changes to the survey-wide photometry include reducing only the optical photometry (F475W, similar to SDSS g; and F814W, similar to I-band), which, for the purposes of this study provide the most information.  The UV imaging is significantly shallower ($\sim$ 2 mag) than the optical, even along the main sequence, resulting in many fewer sources detected.  The IR also yields far fewer main sequence stars as a result of significant crowding due to the lower angular resolution of the WFC3/IR camera.  An image and optical CMD of a typical cluster used in this study are shown in Figures \ref{fig:image} and \ref{fig:cmd}.

In principle, the full spectral energy distribution of each star can increase the degree of precision to which we know its mass, and, in turn provide improved constraints on the IMF.  However, the UV and IR photometry in PHAT clusters typically contains fewer than $\sim$ 50\% the number of stars detected in the optical, reducing their statistical utility for constraining the MF slope \citep[cf.][]{weisz2013a}.  Further, the UV and IR bolometric corrections are not as certain as those in the optical, particularly for massive, metal rich stars, which can introduce systematics into our analysis.  Finally, the modest gain (at best) in precision on the IMF slope of each cluster must be balanced against the computational cost of processing $>$10 million additional artificial star tests (ASTs).  Thus, by only reducing and analyzing the optical data, we retain virtually all statistical leverage on measuring the MF slope, but save considerably on computational costs.  

However, analysis of the full stellar SEDs are valuable for searching for the most massive stars in clusters.  In addition to the slope, the maximum stellar mass plays a fundamental role in defining the high-mass IMF, but is unconstrained by the approach we use to measure the MF slope \citep[e.g.,][]{weisz2013a}.  An investigation of the most massive stars in the PHAT area is the subject on an ongoing study and will be presented in a future paper.

For each cluster, we ran $\sim$50,000 ASTs by adding individual stars of known luminosities to PHAT images of each cluster. The ASTs were uniformly distributed over each cluster's CMD and spatially distributed approximately according to each cluster's light profile. Extensive testing showed that such a spatial scheme was necessary to accurately characterize the completeness and photometric uncertainty profiles of the clusters, as a simple uniform spatial scheme results in overly optimistic completeness and photometric uncertainty profiles.

\subsection{Artificial Clusters}
\label{sec:sample}
 
We use extensive sets of artificial clusters to verify the accuracy of each component of our MF determinations, including photometry, ASTs, and CMD modeling. The artificial clusters were designed to match the physical properties of our real cluster sample and span a range of physical parameters (age, mass, size, extinction) that encompass the real clusters. The artificial clusters were inserted into images at a range of galactocentric radii and environments to capture the full range of background densities in PHAT.  Overall, the artificial cluster tests confirmed that we can accurately recover the true MF from clusters in PHAT imaging, along with all other cluster physical properties.

\section{Methodology}
\label{sec:methodology}

\subsection{Measuring the Stellar Mass Function of a Cluster}
\label{sec:match}

To measure the MF of a cluster we construct synthetic models of its optical CMD using \texttt{MATCH}, a CMD fitting program described in \citet{dolphin2002}.  Briefly, for a given stellar evolution library, binary fraction, IMF, age, metallicity, extinction, and distance, \texttt{MATCH} constructs a simple stellar population (SSP) in the desired filter combinations. This synthetic SSP is then convolved with completeness and photometric biases and uncertainties as determined by the ASTs.  The resulting synthetic CMD is then linearly combined with a suitable CMD of foreground/background populations, with the relative contribution of each controlled by a scale  parameter.  For a given set of these parameters, the synthetic Hess diagram, the binned number density of stars in the CMD, is compared to the observed CMD using a Poisson likelihood function, which is necessary to account for the effects of sparse sampling.  The procedure is repeated over a large grid of parameters in order to map all of likelihood space. For this analysis, we adopted the Padova stellar evolution models \citep{girardi2002, marigo2008, girardi2010} and we have listed the other model parameters and their ranges in Table \ref{tab:matchparams}.

Of these parameters, we limited the metallicity to a fairly small range near \zsun. This was done because (a) the cluster CMDs are too sparsely populated to provide robust metallicity constraints and (b) the current metallicity of M31's disk is known to be approximately solar, with a very weak gradient based on H{\sc II} region  and supergiant abundances \citep[e.g.,][]{venn2000, trundle2002, sanders2012, zurita2012}.

Similarly, we adopted a fixed binary fraction of 0.35, where the mass ratio of the stars is drawn from a uniform distribution.  Historically, stellar multiplicity has been problematic for high-mass IMF determinations from luminosity functions \citep[e.g.,][]{maizapellaniz2005, maizapellaniz2008}.  However, for CMD analysis, it is much less of an issue for two reasons.  First, the addition of color information provides clear separation between the single and binary star sequences, minimizing confusion between the two. In contrast, when only luminosity information is available, the binary fraction and IMF slope are essentially degenerate parameters.  Second, stellar multiplicity is observationally less important for high-mass stars compared to sub-solar mass stars.  For stars above 1 \msun, the majority of companion stars are likely to be significantly less massive and less luminous than the primary \citep[cf.][]{kobulnicky2014}, resulting in minimal change to the CMDs.  

\begin{deluxetable}{cccc}
\tablecaption{Parameters for Measuring the MF of an M31 Star Cluster}
\tablecolumns{4}
\tablehead{
\colhead{Parameter} &
\colhead{Range} &
\colhead{Grid Resolution} &
\colhead{Notes} \\
\colhead{(1)} &
\colhead{(2)} &
\colhead{(3)} &
\colhead{(4)} \\
}

\startdata
$\Gamma$ & (-1.2,5) & 0.2 dex & MF slope for stars $\ga$ 2 \msun \\
Binary Fraction & 0.35 & & Fraction of Stars with a Companion \\
Log(Age)& (6.60, 9.0) & 0.1 dex & Cluster Age \\
A$_V$ & (0.0, 3.0) & 0.05 dex & Line of Sight Extinction \\
(m-M)$_0$ & 24.47 & & Adopted Distance Modulus to M31 \\
log(Z) & (-0.2, 0.1) & 0.1 dex & Metallicity
\label{tab:matchparams}
\tablecomments{Parameters and their ranges used for modeling M31 cluster CMDs.  The adopted distance modulus is from the tip of the red giant branch measurement made by \citet{mcconnachie2005}.}
\end{deluxetable}

\begin{figure}[t!] 
\epsscale{1.2}
 \plotone{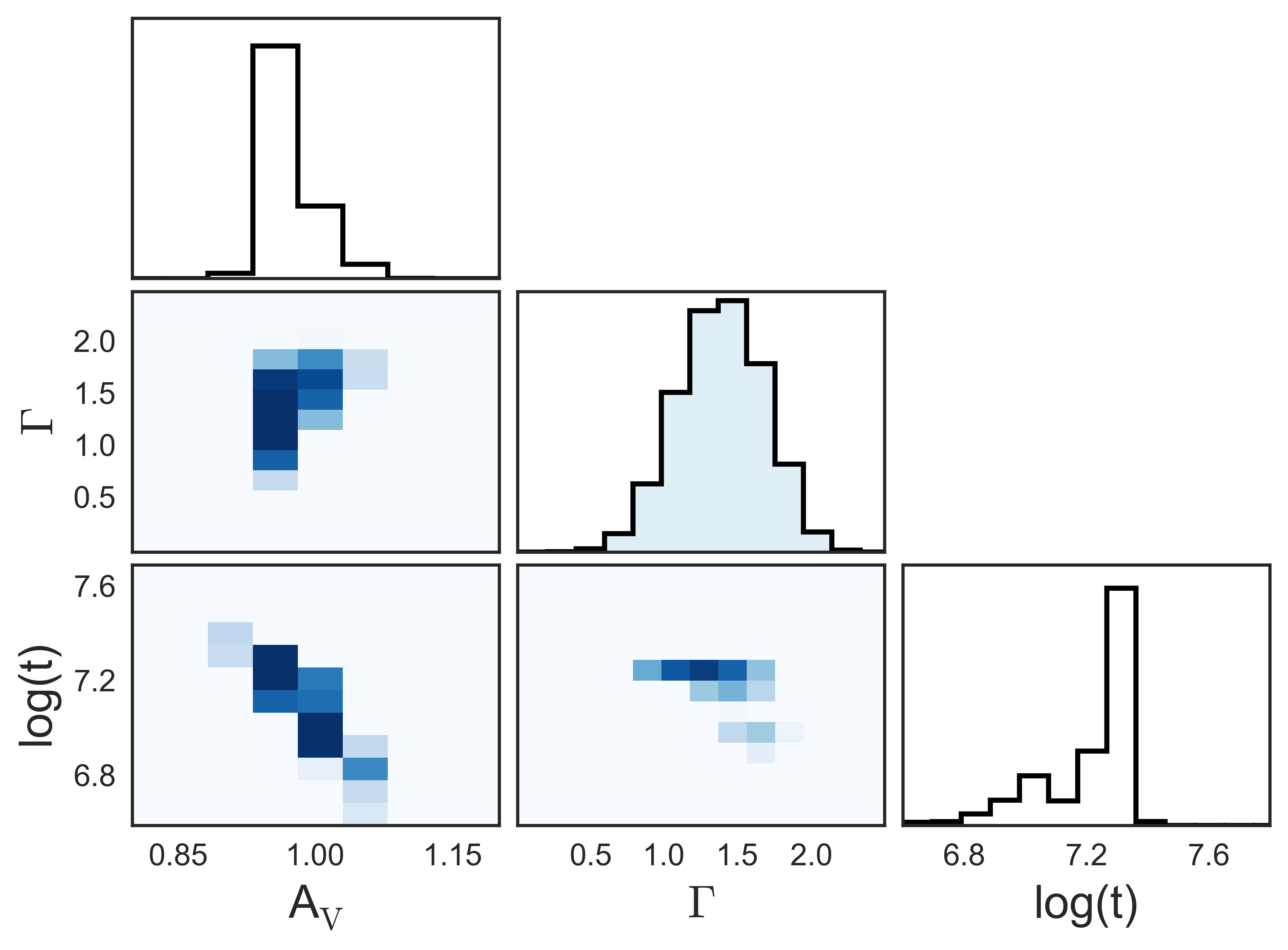} 
\caption{The joint and marginalized distributions of age, MF slope, and extinction for a single young M31 cluster AP 94 that resulted from modeling the cluster's optical CMD, which is shown in Figure \ref{fig:cmd}. The dark points in each panel reflect the regions of high probability, and the pixel sizes reflect the grid resolution as indicated in Table \ref{tab:matchparams}.  We use the marginalized distribution for $\Gamma$ (shaded histogram) in the subsequent modeling of the distribution of MF slopes (\S \ref{sec:inference}). }
\label{fig:triangle_match}
\end{figure}

Within PHAT, we verified that our choice in binary fraction does not change our results.  For example, in the case of AP~94 (Figures \ref{fig:cmd} and \ref{fig:triangle_match}), we modeled the CMD with binary fractions of 0.35 and 0.70 and found differences in the resulting IMF slope constraints to be $<$0.02 dex. Tests including different binary fractions on artificial clusters yielded similarly small changes.  

However, we did find that extreme binary fractions of 0 or 1 resulted in significantly worse CMD fits.  In these two cases, the model either entirely populated the single star sequence (binary fraction = 0) or only sparsely populated it (binary fraction = 1), which substantially reduced the resemblance of the model CMD to the observed CMD.  Thus, for the present work, we adopted a fixed binary fraction of 0.35, which appears to be a reasonable value from LMC cluster studies \citep[e.g.,][]{elson1998}.

To model the background, we construct an empirical CMD from an annulus around the cluster that is ten times  larger than the cluster.  The large area is necessary to sample all parts of the CMD to accurately constrain the background scaling parameter. 

Given a set of cluster model parameters (age, extinction, metallicity, and MF slope) the resulting model CMD, and the empirically derived CMD for fore-/background sources are compared to the observed CMD in a likelihood sense \citep{dolphin2002}. The resulting posterior probabilities, $\mathrm{p}(t,A_V,\Gamma )$ are illustrated in Figure \ref{fig:triangle_match} for a typical cluster in the sample. 
The prior on metallicity tightly brackets solar metallicity and the cluster mass emerges as a normalization factor, rather than a fit parameter.  Finally, the smoothness of the $\mathrm{p}(t,A_V,\Gamma )$ of each cluster allows us to interpolate in order to increase the resolution of our PDFs.

\subsection{Inferring the Intrinsic Distribution of Mass Function Slopes From a Set of Clusters}
\label{sec:inference}

One challenge for past MF studies was how to combine measurements from individual clusters into a broader statement about the MF slope of `clusters in general'.  Typically, the variance weighted mean is used to compute the ``average'' MF from an ensemble \citep[e.g.,][]{scalo1986}.  However, this can potentially be a biased estimator (e.g., assumptions of normally distributed uncertainties), and does not necessarily use all available information (e.g., the probability that an object is actually a cluster).  

Here, we adopt a framework that allows us to infer population wide characteristics
about the \textit{distribution of MF slopes}, given a set of noisy measurements.  This model follows the approach and implementation laid out in \citet{hogg2010} and \citet{foremanmackey2014}.  We briefly describe the model below and we refer the reader to those papers for more detail.

As a simple model for the \emph{distribution} of cluster MFs, we assume that it can be described by a normal distribution, $\mathcal{N}(\Gamma, \sigma_\Gamma)$, with a mean of $\Gamma$ and an intrinsic dispersion of $\sigma_\Gamma$.  We initially assume that their parameters are independent of age, mass, and size of the cluster.  If we assume the probability that each object is a cluster, $\Qk_k$, we can then write down a Gaussian mixture model likelihood function for the MF of clusters, where we simply assume that the MFs of falsely presumed clusters have a different $\mathcal{N}(\Gamma_{false}, \sigma_{\Gamma,false} )$. This mixture model mitigates the impact of erroneously identified clusters, without forcing a binary decision on
which clusters have been identified with sufficient confidence.  

More explicitly, the mixture model can be written as

\begin{eqnarray}
p(\gammamath_k\, |\, \thetamath, \Qk_k)  &=& \Qk_k \, \exp\left(-(\Gamma - \gammamath_k) / (2 \, \sigma_\Gamma^2)\right) \nonumber \\ 
&+&  (1-\Qk_k) \, \exp\left(-(\Gamma_{false} - \gammamath_k) / (2 \, \sigma_{false}^2)\right) \nonumber \, \\
\label{eq:imfmodel}
\end{eqnarray} 

\noindent where $\gammamath_k$ is the marginalized PDF for the MF of the k$^{th}$ object and $\Qk_k$ is the probability that the k$^{th}$ object is a cluster.  $\Gamma$ and $\sigma_\Gamma^2$ are the mean and variance of the Gaussian of interest for the clusters and $\Gamma_{false}$ and $\sigma_{false}^2$ are nuisance parameters for possible contaminating sources. Simply put, with this model, the more probable an object is a cluster the more it contributes to the MF parameters of interest.  In this model context, a universal Kroupa MF would be represented by $\Gamma=1.30$ and $\sigma_{\Gamma}^2\rightarrow0$, and universal Salpeter MF by $\Gamma=1.35$ and $\sigma_{\Gamma}^2\rightarrow0$.

\begin{figure}[t!] 
\epsscale{1.1}
 \plotone{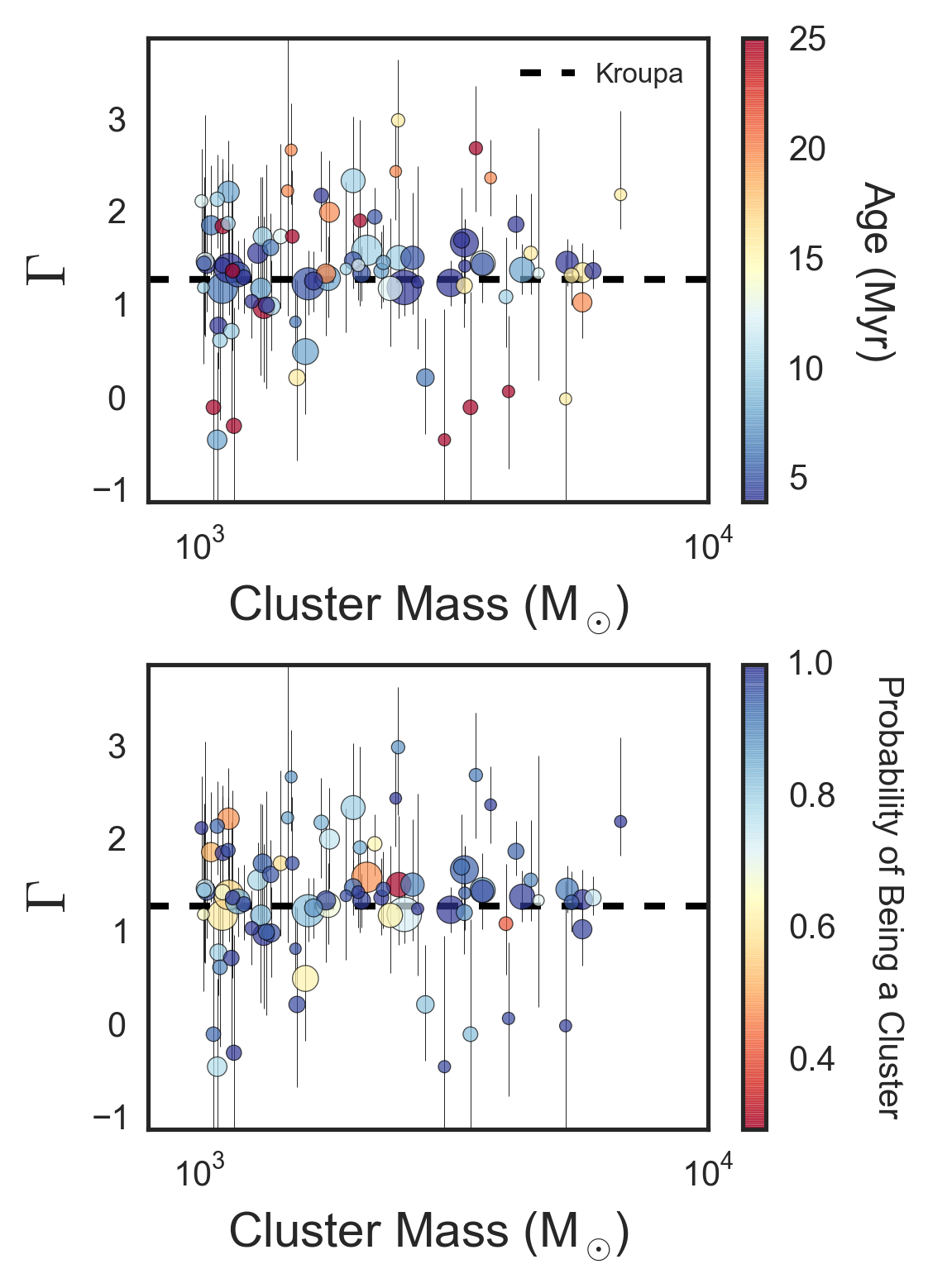} 
\caption{85 young, intermediate mass clusters from PHAT used to measure the high-mass IMF in M31.  Both panels show the median of the probability distribution for for each cluster's MF slope vs. cluster mass.  Points in the top panel are color-coded by their age, while those in the bottom panel by the probability of being a cluster. In all cases, point sizes are proportional to half-light radius and the uncertainties in $\Gamma$ reflect the 68\% confidence interval on the MF slope.}
\label{fig:gammapoints}
\end{figure}

This same framework also readily allows us to explore whether the mean MF slope depends on any of the other cluster properties such as age, mass, and size.  Specifically, we can generalize the model to 

\begin{eqnarray}
\Gamma (t_{k},M_k,r_k )= \overline{\Gamma} + a_m \, {\rm log} \frac{M_k}{M_c} \, + a_t \, {\rm log} \frac{t_{k}}{t_{c}} + a_r \, {\rm log} \frac{r_k}{r_c} \,\,
\\ \nonumber
\label{eq:hypergamma1}
\end{eqnarray}

\noindent where $\{t_k,M_k,r_k\}$ are the most likely age, mass, and size for the $k^{th}$ object and $\{t_c,M_c,r_c\}$ are the mean of our cluster sample in each case. 
We assume priors that are flat in $\overline\Gamma,~\log{\sigma_\Gamma},~a_m,~a_t,~a_r$, 
and sample these parameter's PDFs with the affine invariant ensemble Markov chain Monte Carlo (MCMC) program \texttt{emcee} \citep{foremanmackey2013}\footnote{\url{https://github.com/dfm/emcee}}.

\section{Results}
\label{sec:results}

The analysis laid out in Section \ref{sec:methodology} proceeds in two steps, the first leading to the MF slope measurements for individual clusters, $p(\Gamma_k)$, the second to the characterization of the intrinsic, i.e., error corrected, distribution of mass function slopes in M31 clusters. 

The result of the first step, the marginalized PDF, $\mathrm{p}(\Gamma_k)$, for all 85 sample clusters is
illustrated in Figure \ref{fig:gammapoints}, showing the MF slope and its uncertainty for each cluster versus its mass.  Points in the top panel are color-coded by age and those in the bottom panels by their probability of being a cluster.  In both cases, the point size is proportional to the cluster's half-light radius. Clusters with larger half-light radii have lower crowding and thus more accurate mass function determination.  From a physical standpoint, the size of a cluster may also reflect the environmental conditions in which it was born.

\begin{figure}[t!] 
\epsscale{1.2}
\plotone{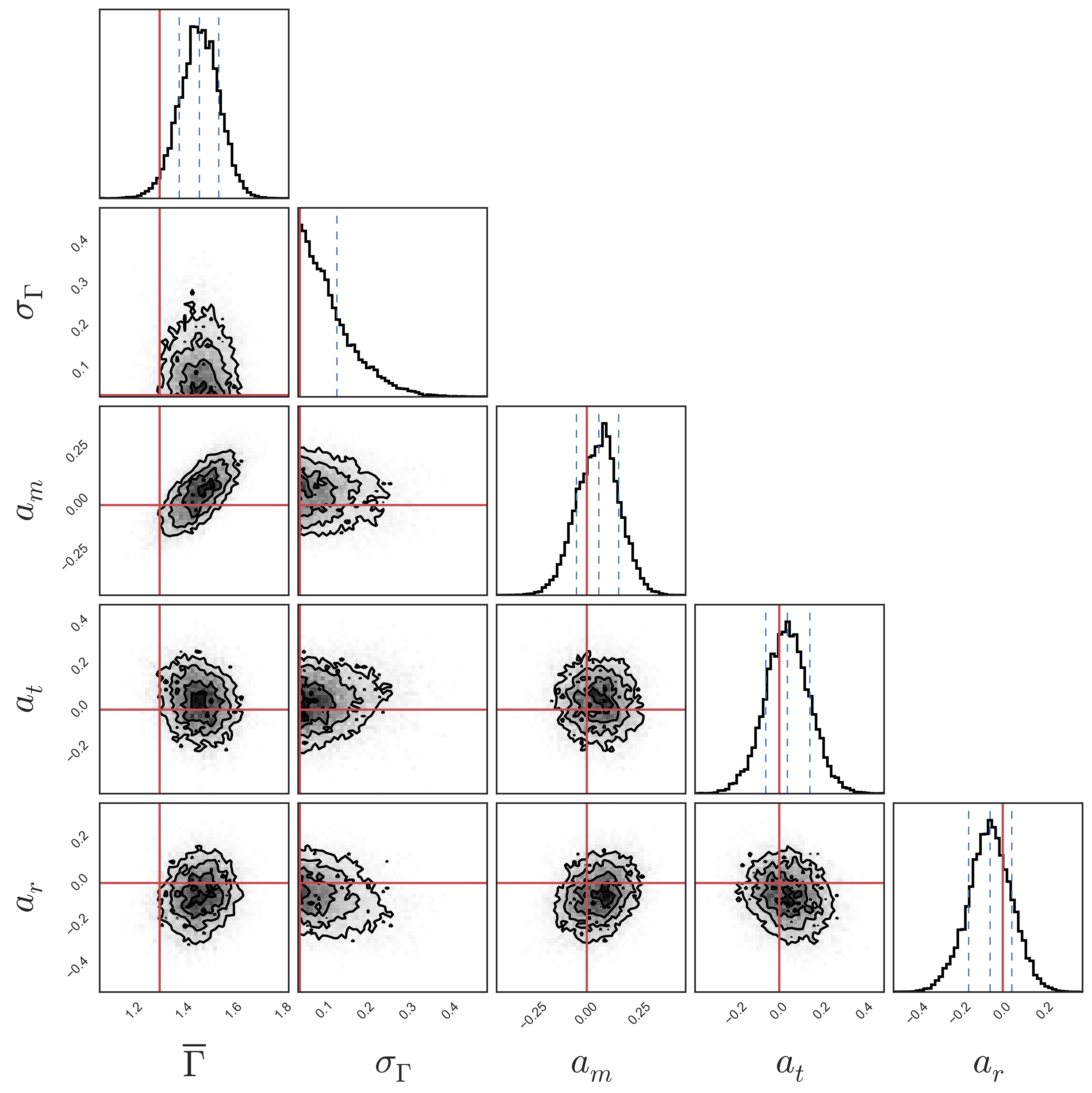} 
\caption{The joint and marginalized distributions for the distribution of cluster MFs described by Equations \ref{eq:imfmodel} and \ref{eq:hypergamma1}. In all panels, except for $\sigma_{\Gamma}$, the dashed blue lines represent the 16th, 50th, and 84th percentiles of each marginalized distribution. The distribution for $\sigma_{\Gamma}$ is highly non-Gaussian, making the median a poor point estimate.  Instead, we use the mode as a point estimate, as it reflects the most probable portion of the PDF, and use the 68th percentile to represent an upper limit. For reference, the solid red lines indicate the value of a Kroupa IMF slope in the $\overline{Gamma}$ panels, the upper uncertainty on $\sigma_{\Gamma}$ for a universal IMF as determined by similar analysis of artificial clusters, and a value of zero in the remaining panels.  This plots shows that the mean IMF slope in M31 clusters is steeper than Kroupa, that the recovered scatter is consistent with expectations of a universal IMF, with a small tail to larger values, and that there are no significant trends between the IMF slope and age, mass, or size of the clusters. }
\label{fig:triangle}
\end{figure}

While we have used a different methodology here than \citet{weisz2013a}, the error bars on the MF are similar in the two approaches, limited in both cases by the number of well-detected cluster members and by the range of stellar masses they span.  For example, consider AP 94, whose CMD and derived physical property PDFs are shown in Figures \ref{fig:cmd} and \ref{fig:triangle_match}.  Effectively, there are $\sim$120 stars (after background correction) between $\sim$ 3 and 11 \msun\ on the upper MS that are used for MF determination.  Given these numbers, from Figure 7 in \citet{weisz2013a}, we expect a theoretical uncertainty of $\sim$$\pm$0.4 on its MF slope.  In comparison, the 1-$\sigma$ uncertainty from the marginalized MF PDF in Figure \ref{fig:triangle_match} is $\pm$0.45, which is very close to theoretical expectation.  Uncertainties in cluster membership, the exact mass range on the main sequence, etc., can explain the small difference in these two values. In general, the cluster MF slopes are generally less precise for older and lower mass clusters, and follow the expected precision relationship derived in \citet{weisz2013a}.

\begin{figure}[t!] 
\epsscale{1.2}
\plotone{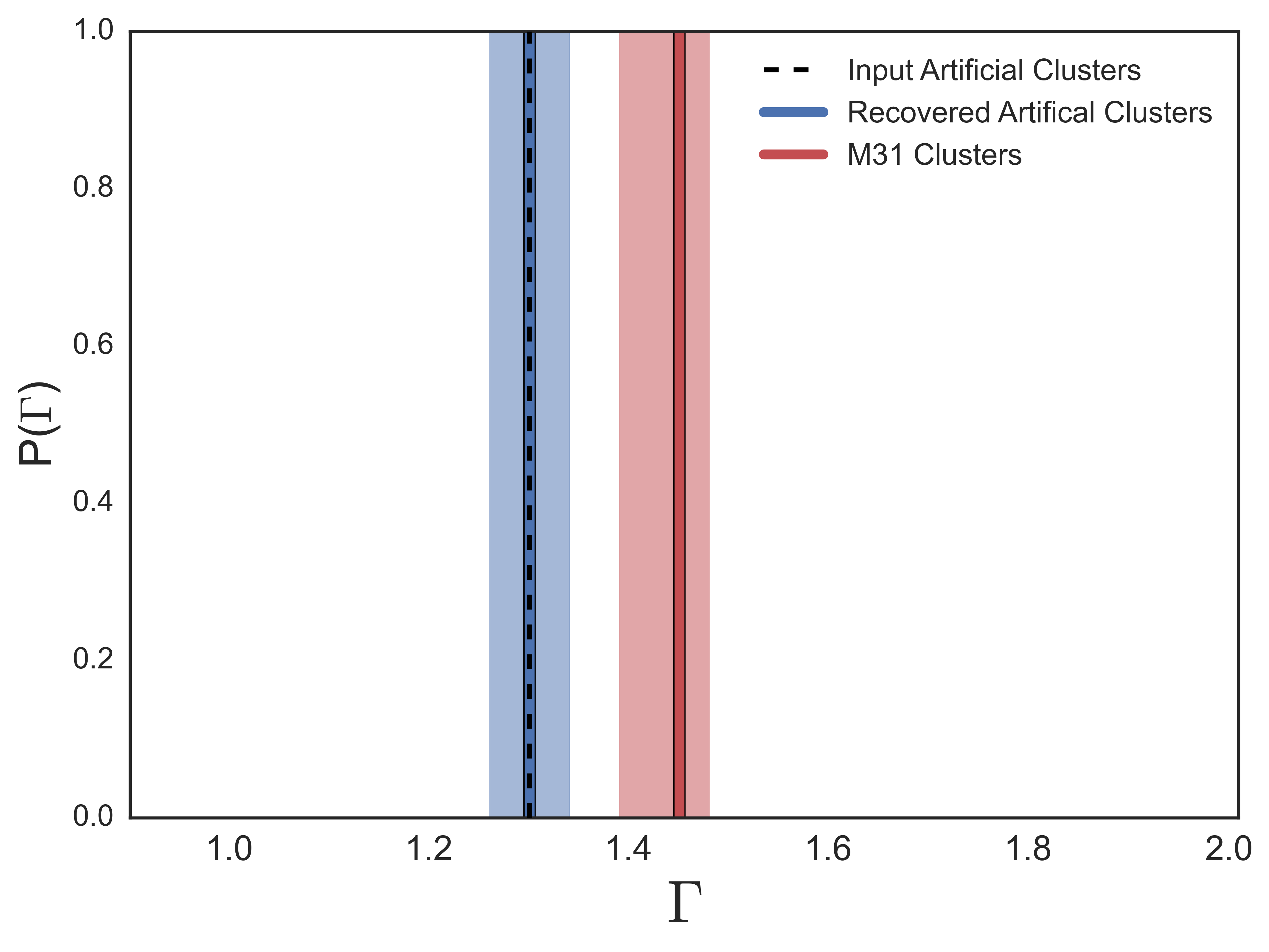} 
\caption{A comparison between the expected IMF slope, $\Gamma$, from artificial clusters inserted into PHAT images (blue) and the real clusters (red), under the assumption that the underlying IMF slope is universal, i.e., $\sigma_{\Gamma}  \rightarrow 0$.  The solid lines indicate the most probable values, while the shaded regions are the 68\% confidence intervals.  The uncertainties on the two measurements begin to overlap at a 2-$\sigma$ level.}
\label{fig:IMFmodel}
\end{figure}

Figure \ref{fig:gammapoints} also gives a qualitative impression of the ensemble properties of MF measurements.  The majority of clusters have MF slopes that are near the Kroupa (1.30) or Salpeter (1.35) slopes; $\sim70\%$ of the clusters MF slopes are within $\sim$ 1-$\sigma$ of Kroupa and $\sim$92\% are within 2-$\sigma$.  Remarkably, in this large sample of young clusters we find that none have extremely steep or flat MF slopes. At least for the parameter space covered by our cluster sample, extreme MF slope outliers ($>3-\sigma$) are quite rare.

The plots also show the lack of strong correlations between cluster MF slope and their physical properties.  From visual inspection, it is clear that the MF slopes do not show significant trends as a function of age, mass, size, or probability of being a cluster.

These impressions from Figure \ref{fig:gammapoints} are quantified in the second step of our analysis, where we propose and constrain 
a simple model for the cluster's distribution of MFs (see \S\ref{sec:inference}), and constrain its parameters through the comparison with
all 85 $\mathrm{p}(\Gamma_k)$.  

Figure \ref{fig:triangle} illustrates the result for the five parameters in our model: 
the mean MF slope, $\Gamma$, and its intrinsic dispersion $\sigma_\Gamma$, and the three coefficients that represent linear trends between the MF slope and cluster age, mass, and size.  This plot shows that the mean MF slope ($\overline{\Gamma}=1.46^{+0.04}_{-0.07}$) is steeper than a Kroupa IMF, that the scatter ($\sigma_{\Gamma}=0.03^{+0.1}_{-0.00}$) is consistent with expectations from a universal IMF as determined by similar analysis of the artificial clusters, and that there are no significant trends between MF slope and cluster physical properties.  The small degree of scatter is particularly interesting, giving the visual impression of large variation between single clusters Figure \ref{fig:gammapoints}.  The typical cluster has an uncertainty on its MF slope of $\sim$ 0.5 dex, but that ensemble scatter scales roughly as $1/\sqrt(N_{\rm clusters})$, indicating that a large and homogenous sample of clusters are needed to statistically identify a universal IMF.  Finally, the limit on $a_t$ implies that there is no evidence that  the MF slope changes between 4 and 25~Myr.  We list summary statistics for each distribution in Table \ref{tab:params} along with the same statistics for our analysis of the population of artificial clusters.

There are two reasons why the artificial clusters results are not an exact delta function. First, the finite grid resolution in p$(\Gamma_k)$ sets a floor of $\sigma\sim0.03$; second, as $\sigma_\Gamma$ is a positive definite quantity, its most likely estimate from noisy data must be significantly positive.   However, this effect is quite small, as the lower limit on the scatter for the artificial clusters is consistent with the minimum possible value, given the resolution of our individual cluster PDFs.

\begin{deluxetable}{cccc}
\tablecaption{IMF Parameters for Ensemble of Clusters}
\tablecolumns{4}
\tablehead{
\colhead{Parameter} &
\colhead{Artificial} &
\colhead{Real} &
\colhead{Real} \\
\colhead{} &
\colhead{Clusters} &
\colhead{Clusters} &
\colhead{Clusters} \\
\colhead{(1)} &
\colhead{(2)} &
\colhead{(3)} &
\colhead{(4)} 
}
\startdata
$\overline{\Gamma_1}$ & $1.29^{+0.05}_{-0.04}$ & $1.46^{+0.04}_{-0.07}$ & $1.45^{+0.03}_{-0.06}$ \\
$\sigma_{\Gamma_1}$ & $0.03^{+0.02}_{-0.00}$ & $0.03^{+0.1}_{-0.00}$ & $\cdots$ \\
$a_m$ & $0.01^{+0.08}_{-0.08}$ & $0.05^{+0.10}_{-0.10}$ & $\cdots$  \\
$a_t$ & $0.05^{+0.07}_{-0.08}$ & $0.06^{+0.12}_{-0.09}$ & $\cdots$  \\
$a_r$ & $-0.02^{+0.10}_{-0.12}$ & $-0.05^{+0.09}_{-0.11}$ & $\cdots$ 
\tablecomments{Column (2) -- Values of the five parameter model for the ensemble IMF from Equations \ref{eq:imfmodel} and \ref{eq:hypergamma1}.  In the case of $\sigma_{\Gamma}$, we quote the mode of its marginalized PDF, along with the upper 68\% confidence interval, as the distribution is highly non-Gaussian. Column (3) -- Value of the IMF slope by fixing all other parameters to zero, i.e., the assumption that the IMF is `universal'.  The listed values reflect the median and surrounding 68\% confidence interval.}
\label{tab:params}
\end{deluxetable}

Given that the scatter is consistent with expectations for a universal IMF, we can consider a model in which all the clusters have the same underlying MF i.e., $\sigma \rightarrow$ 0.    We find an ensemble-wide value of $\Gamma=1.45^{+0.03}_{-0.05}$, which is consistent with results from the hierarchical model.  For comparison, the same analysis applied to the artificial clusters yields a value of $\Gamma=1.30^{+0.04}_{-0.04}$.  Uncertainties on the two mean values begin to overlap at the $\sim$ 2-$\sigma$ level.  We plot the results of the `universal' MF model in Figure \ref{fig:IMFmodel}.

\section{Discussion}
\label{sec:discussion}

The quality of the PHAT data has enabled important progress on three issues: (1) quantifying the extent to which an MF measurement reflects an \emph{initial} MF (IMF) estimate; (2) determining accurately what the mean IMF slope is in typical solar-metallicity (cluster) populations of a large disk galaxy; and (3) assessing the extent to which the IMF is `universal', in the sense that it shows very little scatter in the slope, and no trends of the slope with cluster properties. The fundamental conceptual limitation of the present approach is that it only applies to clusters that remain grouped for longer than 4~Myr, i.e., they have not (yet) disrupted. 
Our conclusions are based on measurements of the high-mass IMF that combine a principled and systematic constraint of each cluster's MF slope,
from CMD analysis, with a probabilistic modeling framework to analyze the distribution of MF slopes.

Overall, we find that the high-mass IMF for clusters in M31 is consistent with a universal IMF.  That is, as discussed in \S \ref{sec:results} and illustrated in Figures \ref{fig:triangle} and \ref{fig:IMFmodel}, the favored high-mass IMF model has a steeper mean slope than Salpeter/Kroupa with a small scatter that is inline with expectations for a universal IMF.  Importantly, we also show that the mean MF slope in clusters of different ages changes by $\Delta\Gamma\la 0.1$ per decade of age.This suggests that the MF slope is unchanged over the first $\sim 25$~Myrs, and provides direct observational evidence that the measured MF represents the IMF.  In this sense, our result represents an important reference point for understanding the potential evolution of the cluster MF with cluster physical parameters (age, mass, size) and local star-forming conditions, both of which will be explored in \citet{weisz2015}.

Of these results, the most robust determination is the mean IMF slope.  Even if we assume that all of the clusters come from the same underlying IMF (i.e., no scatter) the steep slope remains.  The lack of correlations between mean IMF slope and cluster physical properties suggests that other potential sources of bias (e.g., mass segregation) contribute at a negligible level to our measurement of the mean.

\begin{deluxetable}{ccc}
\tablecaption{Select Literature High-Mass IMF Slopes}
\tablecolumns{3}
\tablehead{
\colhead{Cluster} &
\colhead{$\Gamma$} &
\colhead{M$_{\mathrm upper}$} \\
\colhead{} &
\colhead{} &
\colhead{(M$_{\odot}$)}  \\
\colhead{(1)} &
\colhead{(2)} &
\colhead{(3)}  
}
\startdata
& LMC & \\
\hline
LH 6 & 1.7$\pm$0.4 & 85 \\
LH 9 & 1.7$\pm$0.4 & 55 \\
LH 10 & 1.7$\pm$0.4 & 90\\
LH 38 & 1.7$\pm$0.4 & 85 \\
LH 47/48 & 1.3$\pm$0.2 & 50 \\
LH 58 & 1.4$\pm$0.2 & 50 \\
LH 73 & 1.3$\pm$0.4 & 65 \\
LH 83 & 1.3$\pm$0.5 & 50 \\
LH 114 & 1.0$\pm$0.1 & 50 \\
LH 117/118 & 1.6$\pm$0.2 & $>$120  \\
R136 & 1.3$\pm$0.1 & $>$120 \\
\hline
& MW & \\
\hline
NGC 6823 &  1.3$\pm$0.4 & 40 \\
NGC 6871 &  0.9$\pm$0.4 & 40 \\
NGC 6913 &  1.1$\pm$0.6 & 40 \\
Berkeley 86  & 1.7$\pm$0.4 & 40 \\
NGC 7280 &  1.7$\pm$0.3 & 65 \\
Cep OB5 &  2.1$\pm$0.6 & 30 \\
IC 1805 &  1.3$\pm$0.2 & 100 \\
NGC 1893  & 1.6$\pm$0.3 & 65 \\
NGC 2244  & 0.8$\pm$0.3 & 70 \\
NGC 6611  & 0.7$\pm$0.2 & 75 \\
Cyg OB2  & 0.9$\pm$0.2 & 110 \\
Tr 14/16 & 1.0$\pm$0.2 & $>$120 \\
h \& $\chi$ Per & 1.3$\pm$0.2 & 120 
\tablecomments{High-mass IMF slopes from the massive star studies in the MW and LMC clusters as listed in
\citet{massey1998} and \citet{massey2003}.  The dynamic mass range of the clusters extend from $>$1 \msun\ to the mass listed in column (3). }
\label{tab:massey}
\end{deluxetable}

In contrast, it is more challenging to interpret the significance of the upper bound on the scatter.  The results of our inference suggest that a universal IMF is most probable. However, there is a tail to the distribution that extends to $\gtrsim$ 0.1 dex, which exceeds the tail of the same distribution for the artificial clusters.  We believe that this scatter is likely due to sources of uncertainty not included in our modeling that may artificially enhance the scatter relative to expectations from the artificial clusters.  For example, mismatches between the stellar models and real stars are not captured by the artificial cluster tests.  However, the degree to which model-data mismatches contribute to the scatter is hard to quantify.  

Differential effects may also contribute to the observed scatter.  For example, significant cluster-to-cluster variations in mass segregation or binarity could contribute to the scatter in ways not captured by the artificial clusters or our model. Similarly, we impose a single binary fraction for all masses, which may not be accurate, given that the most massive stars may be more likely to have equal mass companions than intermediate mass stars \citep[e.g.,][]{sana2013, kobulnicky2014}.  At the same time, given the modest size of the upper bound, we expect these effects are, at most, fairly small.

Overall, it is remarkable that the IMF slope is consistent with universal in this mass, metallicity, and cluster density regime.  As summarized in \citet{kennicutt1998b}, quantifying the degree of scatter is one of the most critical aspects of understanding the IMF, and ultimately how stars form out of dense gas.  The intrinsic scatter we find suggests that individual clusters should only minimally deviate from the fiducial value, and that extreme outliers (e.g., $\gtrsim 3-\sigma$) are quite rare. The identification of such objects may represent insight into a mode of star-formation that truly differs from what we observe in this regime.

\begin{figure}[tp!] 
\epsscale{1.2}
 \plotone{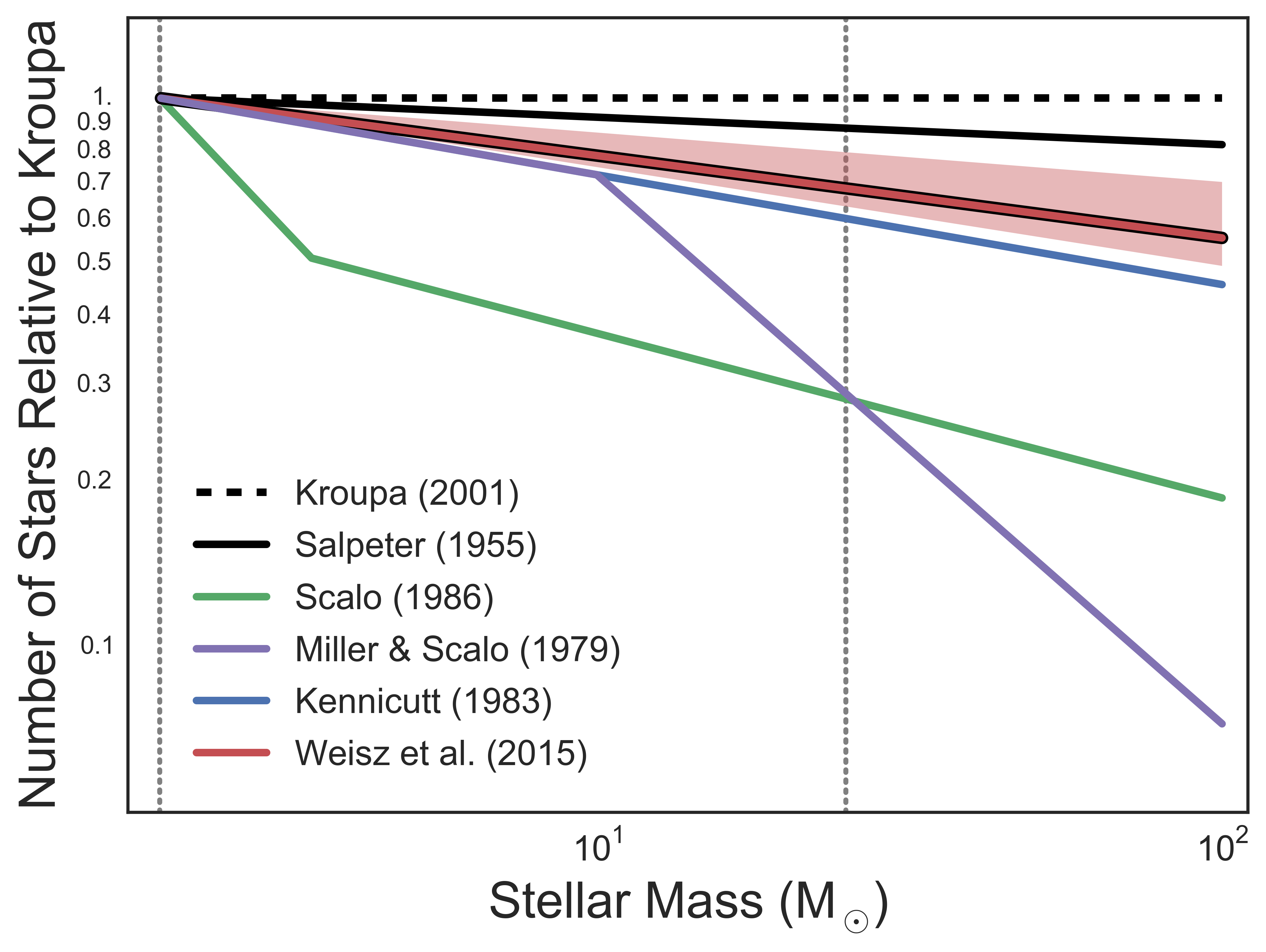} 
\caption{The number of stars predicted by various common IMF models, relative to Kroupa (dashed line). 
The vertical dotted lines indicate the mass range over which the high-mass IMF was measured in M31.  Our median IMF (solid red line) is essentially identical to that of \citet{kennicutt1983} ($\Gamma_{\rm Kennicutt}=1.5$).  However, we also find a small spread, which is indicated by the red band.  Our IMF model indicates a modest deficit of stars $>$2 \msun\ relative to the canonical Kroupa IMF.}
\label{fig:relativeIMF}
\end{figure}

\begin{deluxetable*}{ccccccccccc}
\tablecaption{Number and Fraction of Massive Stars for Different IMFs}
\tablecolumns{11}
\tablehead{
\colhead{IMF} &
\colhead{$N{\star}\ge2 M_{\odot}$} &
\colhead{$N{\star}\ge8 M_{\odot}$} &
\colhead{$N{\star}\ge20 M_{\odot}$} &
\colhead{$N{\star}\ge50 M_{\odot}$} &
\colhead{$N{\star}\ge90 M_{\odot}$} &
\colhead{Frac. $\ge$2 M$_{\odot}$} &
\colhead{Frac. $\ge$8 M$_{\odot}$} &
\colhead{Frac. $\ge$20 M$_{\odot}$} &
\colhead{Frac. $\ge$50 M$_{\odot}$} &
\colhead{Frac. $\ge$90 M$_{\odot}$} \\
\colhead{} &
\colhead{$\frac{10^4 \rm stars}{10^6 \, M_{\odot}}$} &
\colhead{$\frac{10^3 \rm stars}{10^6 \, M_{\odot}}$} &
\colhead{$\frac{10^3 \rm stars}{10^6 \, M_{\odot}}$} &
\colhead{$\frac{10^2 \rm stars}{10^6 \, M_{\odot}}$} &
\colhead{$\frac{10^1 \rm stars}{10^6 \, M_{\odot}}$} &
\colhead{} &
\colhead{} &
\colhead{} &
\colhead{} &
\colhead{} \\
\colhead{(1)} &
\colhead{(2)} &
\colhead{(3)} &
\colhead{(4)} &
\colhead{(5)} &
\colhead{(6)} &
\colhead{(7)} &
\colhead{(8)} &
\colhead{(9)} &
\colhead{(10)} &
\colhead{(11)}  
}
\startdata
Kroupa & 6.9 & $1.1$  & 3.0 & $6.2$ & $6.3$ & 1 & 1 & 1 & 1 & 1  \\
Salpeter & 6.8 & $1.0$  & $2.7$ & $5.4$ & $5.3$ & 0.97 & 0.90 & 0.87 & 0.84 & 0.83 \\
M31 & $6.7^{+0.1}_{-0.0}$ & $0.88^{+0.08}_{-0.04}$  & $2.2^{+0.3}_{-0.2}$  & $4.0^{+0.8}_{-0.4}$ &   $3.8^{+0.9}_{-0.4}$ & $0.90^{+0.04}_{-0.01}$ & $0.74^{+0.09}_{-0.04}$ & $0.66^{+0.11}_{-0.06}$ & $0.60^{+0.13}_{-0.07}$ & $0.56^{+0.14}_{-0.06}$ \\
Kennicutt & 6.7 & $0.82$  & $1.9$ & $3.5$ & $3.2$ & 0.87 & 0.67 & 0.57 & 0.50 & 0.46 \\
Miller \& Scalo & 6.8 & $0.66$  & $0.82$ & $0.81$ & $0.56$ & 0.85 & 0.51 & 0.23 & 0.11 & 0.08 \\
Scalo & 5.6 & $0.20$  & $0.43$ & $0.72$ & $0.65$ & 0.54 & 0.14 & 0.10 & 0.08 & 0.07

\enddata
\tablecomments{The number of expected stars above the specified mass limit per $10^6$ \msun.  For these calculations, we have adopted a \citet{kroupa2001} IMF below 2 \msun, the specified IMF above 2\msun, and assumed lower and upper mass bounds for the total IMF to be 0.08 \msun\ and 100 \msun.  Columns (7) - (11) give the fraction of stars formed above the listed mass, relative to the standard Kroupa IMF.}
\label{fig:nstars}
\end{deluxetable*}

\subsection{Comparison with Other High-Mass IMF Slopes in the Local Group}
\label{sec:LG}

It is important to place our results into context of other high-mass IMF studies in the Local Group (LG).  Among the strongest evidence for the universality of the high-mass IMF in the LG comes from a series of papers that combine spectroscopy and photometry of resolved, massive stars in clusters and OB associations in the LMC and MW \citep{massey1995, massey1995a, massey1998, massey2003}.  

To summarize, these studies found (a) no drastic differences between the IMF slopes of very young, $<$ 5 Myr star forming regions in the MW and LMC, despite large differences in metallicity; and (b) that the resulting IMF slopes are broadly consistent with Salpeter/Kroupa.  Combined, these studies represented a major milestone in our understanding of the high-mass IMF and its (in-)sensitivity to environment.  

Although re-analyzing these studies is beyond the scope of this paper, we can quantify what the ensemble high-mass IMF slopes are in the LMC and MW using data from \citet{massey2003}.  The purpose is to give a general sense of how different our M31 result is from other LG environments.  To perform this analysis, we extracted the necessary data from \citet{massey2003} and have listed it in Table \ref{tab:massey}.  We simply use the model from Equation \ref{eq:imfmodel} with $\Qk_k=1$ for each object, i.e., we know they are all \textit{bone fide} clusters, to infer the mean and variance for the MW and LMC clusters.  We do not model possible age, mass, size dependancies.

From this exercise we find mean values of $\Gamma_{\mathrm MW}=1.16^{+0.12}_{-0.10}$ and $\Gamma_{\mathrm LMC}=1.29^{+0.11}_{-0.11}$.  In both cases, $\sigma_{\Gamma} \sim 0.3-0.4$. The large scatter in the posterior distributions is likely a reflection of underestimated uncertainties and/or systematics, as opposed to true physical variation.

\citet{massey2003} emphasizes that the reported uncertainties are only statistical in nature.  However, even these are likely under-estimated, as discussed in \citet{weisz2013a}. Unfortunately, not all of these literature studies list both the number of stars and mass ranges needed to estimate degree of under-estimation of the random uncertainties. There are also issues of systematic uncertainties in stellar models that can significantly change the masses of the most massive stars used in these studies \citep[e.g.,][]{massey2011}.

However, even from this cursory exercise, we see that the MW, LMC, and M31 do not appear to have\emph{ drastically} different IMF slopes.  While this is quite remarkable given the diversity of environments, it should be tempered by the heterogenous nature of this comparison.  If high-mass IMF variations are subtle, then homogenous and principled analyses are absolutely necessary to uncover them.  At present, only our analysis in M31 is the result of a systematic study of the high-mass IMF over a large number of young clusters.  Any comparison of the high-mass IMF slopes of LG galaxies must come from similarly homogenous data and principled analysis in order to minimize the systematics that have been persistent in IMF studies for the better part of six decades.

\subsection{Practical Usage of High-Mass IMF}
\label{sec:recommend}

Given the analysis presented in this paper and comparison with LG literature, we do not find strong evidence that a universal Salpeter/Kroupa IMF is the best representation of the high-mass IMF slope in the LG.  Although we only rule it out at a $\sim 2-\sigma$ level, there appears to be little basis for this canonical value being correct in the first place\footnote{\citet{salpeter1955} has had remarkable staying power, despite the fact 
that the power-law index of $\Gamma=1.35$
was derived assuming the MW was 6 Gyr old. A more realistic
13 Gyr age of the MW yields a value of $\Gamma=1.05$, which is considered an 
extreme outlier by modern standards \citep[e.g.,][]{salpeter2005, zinnecker2005}.}.  Instead, for practical purposes the high-mass IMF can be represented by a single-sloped power-law with $1.45^{+0.03}_{-0.06}$.  Although this is only 0.1-0.15 dex steeper than Salpeter/Kroupa it does make a difference in the number of high-mass stars formed, and related quantities such as commonly used star formation rate (SFR) indicators.  We discuss these implications further in \S \ref{sec:implications}.   

For general usage, we recommend adoption of the following forms of the the IMF over all stellar masses.  With the M31 high-mass IMF, a modified Chabrier IMF \citet{chabrier2003} has the form

\begin{eqnarray}
\xi(m)\, \Delta(m)\, &=& \frac{0.086}{m} \, exp^{\frac{-(log(m) - log(0.22))^2} {2 \times 0.57^2}} \nonumber \\
\nonumber \\
\xi(m)\, = c\, m^{- (\Gamma+1)}\, , \; \Gamma &=&1.45^{+0.03}_{-0.06} \, ,  \mbox{for }  1.0 < m < 100 \ \, M_{\odot}\, ,
\label{eqn:newchabrier}
\end{eqnarray}

\noindent and a modified \citet{kroupa2001} IMF has the form

\begin{equation} 
\xi(m)\, = c\, m^{- ( \Gamma+1)} \begin{cases} \Gamma = 0.3  &\mbox{for }  0.08 < m < 0.5 \, M_{\odot} \\ 
\Gamma = 1.3  &\mbox{for }  0.5 < m < 1.0 \, M_{\odot} \\
\Gamma = 1.45^{+0.03}_{-0.06} \, ,  &\mbox{for }  1.0 < m < 100 \, M_{\odot} \, .\end{cases}
\label{eqn:newkroupa}
\end{equation}

Here, we have made two extrapolations to match conventional IMF definitions in the literature.  First, we have extrapolated the M31 IMF down to 1 \msun.  For most practical purposes, this extension from 2 \msun\ to 1 \msun\ affects the resulting stellar mass spectra at the level of a few percent.

Second, we have extrapolated the M31 IMF up to 100 \msun.  While our data do not provide good constraints above $\sim$ 25 \msun, there is little other information to go on for mass spectrum of the most massive stars.  Thus, as is commonplace in the literature, we suggest extrapolating the M31 IMF up to the highest masses, but recognize that systematic spectroscopic searches for the massive stars are absolutely critical for understanding the IMF of the most massive stars \citep[e.g.,][]{massey1995a, massey2003}.

Finally, we have assumed that the sub-solar IMF in M31 is similar to that of the Galaxy.  Although not confirmed by direct star counts, \citet{conroy2012} have shown that spectral features sensitive to the low-mass IMF yield a result that is similar to the low-mass Galactic IMF.

\subsection{Broader Implications}
\label{sec:implications}

In Figure \ref{fig:relativeIMF} we illustrate the mass spectra of high-mass stars predicted by various commonly used IMF models.  The plot shows the number of stars expected relative to a Kroupa IMF ($\Gamma=1.3$) as a function of mass. For fair comparison, the slopes have all been normalized at 2 \msun, the lower main sequence mass limit of the PHAT data, and extended to 100 \msun\ on the upper end.  We chose this normalization, as opposed to 1 \msun\ to discuss out ability to distinguish different literature forms of the IMF over the dynamic mass range of the data, which is indicates by the vertical dashed lines. 

The M31 high-mass IMF predicts the formation of fewer stars than Kroupa at all masses.  The red solid line presents our median IMF model and shows that the fractional deficit varies from $\sim$0.9 at $\sim$ 10 \msun\ to $\sim0.7$ at 100 \msun.  The red shaded region reflects the 68\% confidence interval in the number of predicted high-mass stars due to uncertainty on the mean high-mass IMF slope in M31.  At most masses, the uncertainty deviates from the median by $\pm$$\sim$5\%, but increases to  $\pm$15\% for stars $>$ 50 \msun.

The M31 high-mass IMF is similar to other high-mass IMF models in the literature.  It is particularly close to the Kennicutt IMF \citep[$\Gamma=1.5$;][]{kennicutt1983}.  It is also the same as the IMF of \citet{miller1979} over the 1-10 \msun\ range.  The steepness of the \citet{miller1979} IMF above 10 \msun\ is known to be too extreme \citep[e.g.,][]{kennicutt1983, kennicutt1994}.  Further, the high-mass IMF from \citet{scalo1986} has a similar slope ($\Gamma=1.6$) above $\sim$ 4 \msun, but a much steeper slope below it.  Finally, the M31 high-mass IMF predicts fewer high-mass stars than a Salpter IMF for masses $>$ 2 \msun.

We further quantify differences in these IMF models in Table \ref{fig:nstars}.  Here, we have computed the expected number of stars per 10$^6$ \msun\ formed at selected stellar masses assuming a \citet{kroupa2001} IMF below 2 \msun\ and the listed IMFs above 2 \msun.  This table solidifies many of the above points regarding differences in commonly used high-mass IMF models.  It also highlights an important implication for core collapse supernovae, whose progenitors are believed to have masses $\gtrsim$ 8\msun \citep[e.g.,][]{smartt2009}.  For example, the median M31 high-mass IMF model predicts $\sim$25\% fewer stars with masses $\gtrsim$ 8\msun\ compared to a Kroupa IMF.

\begin{deluxetable}{ccccc}
\tablecaption{Updated Broadband SFR Indicators}
\tablecolumns{5}
\tablehead{
\colhead{Band} &
\colhead{$L_{x}$ Units} &
\colhead{log($C_x$)} &
\colhead{log($C_x$)} &
\colhead{$\dot M_{\star}(M31) / \dot M_{\star}$(Kroupa)} \\
\colhead{} &
\colhead{} &
\colhead{(Kroupa)} &
\colhead{(M31)} &
\colhead{} \\
\colhead{(1)} &
\colhead{(2)} &
\colhead{(3)} &
\colhead{(4)} &
\colhead{(5)} 
}
\startdata
FUV &  erg s$^{-1}$ ($\nu L_{\nu}$) & 43.29 & 43.18$^{+0.05}_{-0.02}$ & 1.28$^{+0.06}_{-0.12}$ \\
NUV &  erg s$^{-1}$ ($\nu L_{\nu}$) & 43.14 & 43.04$^{+0.03}_{-0.02}$ & 1.24$^{+0.06}_{-0.11}$ \\
\halpha\ &  erg s$^{-1}$  & 40.91 & 40.73$^{+0.04}_{-0.07}$ & 1.52$^{+0.13}_{-0.24}$ 
\enddata
\tablecomments{A comparison of commonly used SFR indicators for a Kroupa and M31 IMF values.  The values of $L_x$, $C_x$, and $\dot M_{\star}$ follow \citet{kennicutt2012} and Equation \ref{eqn:sfr}.}
\label{tab:sfrindicators}
\end{deluxetable}

Finally, we consider the implications for commonly used SFR indicators \citep[e.g.,][]{kennicutt1998, kennicutt2012}.  Using the Flexible Stellar Population Synthesis code \citep[FSPS;][]{conroy2009, conroy2010a}, we compute the luminosity to SFR conversion coefficients for GALEX far- and near-UV luminosities, along with \halpha, which we list in Table \ref{tab:sfrindicators}.  To do this, we have assumed a constant SFH over the last 1 Gyr, the indicated IMF, mass limits from 0.08 to 100 \msun, and case B recombination to convert the number of ionizing photons to \halpha\ luminosity.  From \citet{kennicutt2012} the canonical conversion from luminosity to SFR can be written as

\begin{equation}
{\rm log}\,  \dot M_{\star}\, ({\rm M}_{\odot} \, {\rm yr}^{-1}) \,=\, {\rm log}\, L_x\, -\, \rm{log}\, C_x
\label{eqn:sfr}
\end{equation}

\noindent where $\dot M_{\star}$ is the SFR, $L_x$ is the luminosity over the bandpass $x$, and $C_x$ is the conversion constant over the same wavelength range.  We have provided updated values for $C_x$ in Table \ref{tab:sfrindicators}. 

From column (5) of Table \ref{tab:sfrindicators}, we see modest differences in the SFR indicators due to an M31 high-mass IMF.  The variations are such that for a fixed luminosity, using the M31 high-mass IMF results in SFRs that are a factor $\sim$1.3--1.5 larger.

\acknowledgments 

The authors would like to thank Nate Bastian, Charlie Conroy, Rob Kennicutt, and John Saclo for their useful comments on this paper and discussions about the IMF and its history.  Support for this work was provided by NASA through grant number HST GO-12055 from the Space Telescope Science Institute, which is operated by AURA, Inc., under NASA contract NAS5-26555. 
 DRW is supported by NASA through Hubble Fellowship grant HST-HF-51331.01 awarded by the
Space Telescope Science Institute. 
DRW also wishes to thank the MPIA and KITP for their generous hospitality during the writing of this paper.  
This work used the Extreme Science and Engineering Discovery Environment (XSEDE), which is supported by National Science Foundation grant number ACI-1053575, was supported in part by the National Science Foundation under Grant No. NSF PHY11-25915, and made extensive use of NASA's
Astrophysics Data System Bibliographic Services. 
 In large part, analysis and plots presented in this paper utilized IPython and
packages from NumPy, SciPy, and Matplotlib \citep[][]{hunter2007,
oliphant2007, perez2007, astropy2013}.

\bibliography{ms.bbl}

\end{document}